\documentclass[12pt,letterpaper]{article}

\usepackage{graphicx,array}
\usepackage{url}
\usepackage{color}
\usepackage{latexsym}
\usepackage{amsthm}
\usepackage{amsmath}
\usepackage{amssymb}
\usepackage{amsfonts}
\usepackage[numbers,sort&compress]{natbib}
\usepackage{bm}
\usepackage{slashed}
\usepackage{mathrsfs}
\usepackage{enumerate}
\usepackage{tikz}
\usepackage{siunitx}
\usepackage{mdframed}
\usepackage{setspace}  
\usepackage{esvect}
\usepackage{esint}
\usepackage{subcaption}
\usepackage[normalem]{ulem}
\usepackage{setspace}

\usepackage[utf8x]{inputenc}%
\usepackage{tcolorbox}%

\usepackage{hyperref} 

\hypersetup{
    colorlinks=true,       
    linkcolor=red,          
    citecolor=blue,        
    filecolor=magenta,      
    urlcolor=blue           
}
\usepackage[all]{hypcap} 

\usepackage{natbib}
\newcommand{\YB}[1]{{\color{blue} #1}}

\newcommand{\nc}{\newcommand}

\nc{\beq}{\begin{equation}}  
\nc{\eeq}{\end{equation}}  
\nc{\beqa}{\begin{eqnarray}}  
\nc{\eeqa}{\end{eqnarray}}  
\nc{\bit}{\begin{itemize}}  
\nc{\eit}{\end{itemize}}

\usepackage{floatrow}
\newfloatcommand{capbtabbox}{table}[][\FBwidth]

\usepackage{blindtext}

\title{ \bf
Muon $\bm{g} - \bm{2}$ in Lepton Portal Dark Matter
 }
 \author{\large Yang Bai$^{\,\star}$ and Joshua Berger$\,^\dagger$}
\date{\small \it 
$^\star$Department of Physics, University of Wisconsin-Madison, Madison, WI 53706, USA\\
$^\dagger$Department of Physics, Colorado State University, Fort Collins, Colorado 80523, USA \\
}

\begin{document}

\maketitle

\setlength{\parskip}{0.2ex}

\begin{abstract}
The Lepton Portal Dark Matter model, in which dark matter states only coupling to the charged leptons, can explain the excess of the muon anomalous magnetic moment measured by the Muon $g - 2$ experiment. In this paper, we demonstrate that real, charge-neutral scalar dark matter with a large number of internal degrees of freedom and a mass approximately degenerate with the charged fermionic mediator state can accommodate the $(g - 2)_\mu$ excess. The model remains consistent with the dark matter relic abundance, direct detection, and indirect detection constraints. The dark matter and its charged fermion partner masses are constrained to be below around 200 GeV. The high-luminosity LHC and future lepton colliders, as well as indirect searches at CTA and GAMMA-400, can test this scenario. 
\end{abstract}

\thispagestyle{empty}  
\newpage  
  
\setcounter{page}{1}  

\begingroup
\hypersetup{linkcolor=black}
\endgroup


\section{Introduction}
\label{sec:intro}

The muon anomalous magnetic moment a key probe of new physics beyond the Standard Model (SM). The latest theoretical calculations have the SM prediction for $a_\mu =(g-2)_\mu/2$  as~\cite{Aoyama:2020ynm}
\beqa
a_\mu^{\rm SM} = 116591810(43) \times 10^{-11} ~, 
\eeqa
with the uncertainty dominated by the hadronic vacuum polarization. On the experimental side, the E821 experiment at the Brookhaven National Laboratory (BNL) has measured this quantity to be~\cite{Bennett:2006fi}
\beqa
a_\mu^{\rm E821}  = 116592089(63) \times 10^{-11} ~. 
\eeqa
This measurement was recently updated by the Muon $g-2$ experiment at Fermilab, which observed~\cite{PhysRevLett.126.141801}
\beqa
a_\mu^{\rm FNAL}  =  116592040(54) \times 10^{-11} ~. 
\eeqa
The combination of these two experimental results yields a measured value of~\cite{PhysRevLett.126.141801}
\beqa
a_\mu^{\rm exp}  =  116592061(41) \times 10^{-11} ~, 
\eeqa
which exceeds the SM prediction by 
\beqa
\Delta a_\mu := a_\mu^{\rm exp}  - a_\mu^{\rm SM} = 251(59) \, \times 10^{-11} ~,
\eeqa
corresponding $4.2 \sigma$ discrepancy. While this result awaits further scrutiny, if confirmed, it is the first discovery of new physics beyond the SM at directly accessible energies that cannot be decoupled and therefore requires a viable explanation.

There are many such possibilities to explain the muon $g - 2$ result. From an effective field theory point of view, we write new physics contributions to the magnetic moment of the muon as a dimension-five effective operator
\beqa
\mathcal{L} \supset  \,\epsilon\,\frac{e\,m_\mu}{16\pi^2 \, \Lambda^2}\, \overline{\mu}\, \sigma^{\mu\nu} \, \mu \, F_{\mu \nu}  ~, 
\eeqa
with $\epsilon=\pm 1$ depending on new physics. This operator contributes to $\Delta a_\mu$ as 
\beqa
\Delta a_\mu = \epsilon\, \frac{e\,m_\mu^2}{4\pi^2 \, \Lambda^2} \approx (251\times 10^{-11})\times \left(\frac{\epsilon}{+1}\right) \left(\frac{183\,\mbox{GeV}}{\Lambda}\right)^2 ~,
\eeqa
which indicates that the muon $g - 2$ excess points to new charged particles in the few hundred GeV mass range. 

As the scale of this operator is within reach of direct experimental probe, we turn to studying its origin in a renormalizable theory. One notable possibility is in the framework of the well-studied Minimal Supersymmetric Standard Model (MSSM)~\cite{Chung:2003fi,Moroi:1995yh,Martin:2001st}. Given the long-standing puzzle of the identity of dark matter, another possibility that there is a correlation between muon $g - 2$ and dark matter states. Along this direction, we consider a class of models known as ``Lepton Portal Dark Matter" (LPDM)~\cite{Bai:2014osa,Chang:2014tea,Agrawal:2014ufa}, which represents a class of simplified dark matter models with dark matter coupling directly with a mediator and a charged lepton. Depending on the spin of the dark matter state, the dark matter sector contribution to $\Delta a_\mu$ could be either positive or negative. To explain the observed positive value of $\Delta a_\mu$, one could have the dark matter to be a scalar and the charged mediator to be a fermion, which is the focus of our work. Even for the minimal model presented in Ref.~\cite{Bai:2014osa}, various experimental searches for dark matter together with a theoretical perturbativity requirement have highly constrained the  $\Delta a_\mu$ preferred parameter space (see Ref.~\cite{Kawamura:2020qxo} for updated status). One notable constraint comes from dark matter indirect detection by searching gamma ray lines from the three-body annihilation with an internal bremsstrahlung photon~\cite{Ackermann:2013uma,Garny:2013ama}. In this paper, we will show that those indirect constraints can easily be relaxed if the dark matter state $X$ has multiple degrees of freedom $n_f$. For large $n_f$, holding the dark matter mass fixed, the required coupling to explain $\Delta a_\mu$ is reduced by a factor of $1/\sqrt{n_f}$ such that the dark matter annihilation cross section is reduced by a factor of $1/n_f^2$. To further suppress various experimental constraints, we will choose the dark matter to be a real scalar that is a fundamental under an $S_{n_f}$ symmetry. 

Our paper is organized as follows. In Section~\ref{sec:model}, we present the LPDM model with particles charged under gauge and global symmetries. In Section~\ref{sec:collider}, we work out the constraints on the model parameter space from various collider searches including the soft lepton searches aiming for degenerate spectra. We then present the formula to explain $\Delta a_\mu$ in this model in Section~\ref{sec:muon-g-2}. Section~\ref{sec:dark-matter} contains the dark matter phenomenology including the thermal abundance in~\ref{sec:thermal}, direct detection in~\ref{sec:direct} and indirect detection in~\ref{sec:indirect}. We conclude our paper in Section~\ref{sec:discussion}. 

\section{The Model}
\label{sec:model}

We consider the SM plus three right-handed neutrinos that provide a Seesaw mechanism to explain the light neutrino mass. In the lepton sector, we have 
\beqa
\mathcal{L} \supset - \, Y_e\, \overline{L}_L\, H \, e_R \,-\,  Y_\nu\, \overline{L}_L\, \widetilde{H} \, \nu_R \,  - \frac{1}{2}\, Y_R\, M_R\, \nu_R^T \,\mathcal{C}\, v_R \,+\, h. c.  , 
\eeqa
with $Y_e$, $Y_\nu$, and $Y_R$ as $3\times3$ mass matrices. Here, $\mathcal{C}$ is the charge-conjugation operator and $\widetilde{H} = i\sigma_2 \, H^*$; $M_R$ sets the general scale for the right-handed neutrinos. Treating the Yukawa matrices as spurion fields, the global symmetry in the lepton sector is $SU(3)_{L_L} \times SU(3)_{e_R} \times SU(3)_{\nu_R}\times U(1)_L$, where we only keep tracking the common lepton number symmetry $U(1)_L$. Under the global symmetry, one has 
\beqa
Y_e \in (3, \overline{3}, 1)_0 \,, \qquad Y_\nu \in (3,1,  \overline{3})_0 \,, \qquad   Y_R \in (1, 1, \overline{6})_{-2} \,. 
\eeqa

To simplify our discussion, we consider a specific structure for $Y_R = y_R\,\mathbb{I}_3$, which breaks $SU(3)_{\nu_R} \rightarrow SO(3)_{\nu_R}$ as well as $U(1)_L$. With this assumption, the light neutrino mass matrix is
\beqa
M_\nu = - \frac{v^2}{2} Y_\nu \, (Y_R\,M_R)^{-1}\, Y_\nu^T  = - \frac{v^2}{2\,y_R\,M_R}\,  Y_\nu\, Y_\nu^T ~,
\eeqa
with $v=246$~GeV as the electroweak scale. In the basis of diagonal charged-lepton mass matrix, one has $M_\nu = U_{\rm PMNS} \, \hat{m}_\nu \, U_{\rm PMNS}^{\rm T}$ with $\hat{m}_\nu$ as the diagonal neutrino mass matrix. 
%
%

In the dark matter sector, we introduce (real) scalars $X^a$ with $a=1,\cdots n_f$ as the dark matter flavor index that contains the dark matter candidate and a vector-like charged fermion $\psi_L$ and $\psi_R$ that are $(1, 1)_{-1}$ under the SM gauge group and have the same gauge charge as $e_R$. There is another global discrete symmetry $S_{n_f}$ associated with the number of species of $X^a$. Under the flavor symmetry, $SU(3)_{L_L} \times SU(3)_{e_R} \times SU(3)_{\nu_R}\times S_{n_f} \times U(1)_L$, 
we have 
\beqa
X \in (1, 1, 1, n_f)_0 \,, \qquad \psi \in (1, 3, 1, 1)_1 ~,
\eeqa
and both are $\mathcal{Z}_2$-odd under a dark parity. The renormalizable interactions are 
\beqa
\mathcal{L} &\supset&  - \,Y^a_X\, X^a\, \overline{\psi^i_L} \, e^i_R \,- \, Y_\psi\, \mu\, \overline{\psi^i_L}\, \psi^i_R \,+\, h.c.  - \frac{1}{2}\, M_X^2\,X^aX^a \nonumber \\
&=&  - \,\lambda\, X^a\, \overline{\psi^i_L} \, e^i_R \,- \, m_\psi\, \overline{\psi^i_L}\, \psi^i_R \,+\, h.c.  - \frac{1}{2}\, M_X^2\,X^aX^a   ~,
\eeqa
with $Y^a_X  = \lambda$ for all $a$ to conserve the $S_{n_f}$ symmetry. In our study, we will consider degenerate masses for all $n_f$ $X^a$ or the dark matter candidate $X$ contains a number of degrees of freedom $n_f$.  For the three electrically-charged fermions,  $\psi^i$ or $\psi_{e, \mu, \tau}$, we assume that that they have an approximately degenerate spectrum, although the flavor symmetry breaking spurion could change the story. For instance, the $\psi_\tau$ field that couples to the $\tau$ lepton could be slightly heavier than the other two $\psi_\mu$ and $\psi_e$.  So, there are totally 5 parameters in our model: $\lambda, M_X, m_\psi, n_f$ and $\Delta m_{\psi_\tau} \equiv m_{\psi_\tau} - m_{\psi_{e, \mu}}$. For most of our studies later, the last parameter $\Delta m_{\psi_\tau}$ does not play a significant role. For the presentation purpose, we also define $\Delta m \equiv m_\psi - M_X$.

To explain $\Delta a_\mu$,  the coupling strength $\lambda$ is required to be large. So, when we consider the renormalization group equation (RGE) running of $\lambda$, we can approximately ignore the SM gauge, Yukawa and quartic couplings. At two-loop level, one has~\cite{Staub:2013tta}
\beqa
\frac{d\lambda}{d\ln{\mu}} \equiv \beta_\lambda(\lambda) = 5\,n_f\, \frac{\lambda^3}{(4\pi)^2} - \frac{57}{4}\,n_f^2\, \frac{\lambda^5}{(4\pi)^4} ~.
\eeqa
Note that there exists an ultra-violet (UV) fixed point value $\lambda_*$ with $\beta(\lambda_*) = 0$ and 
\beqa
\lambda_* = \frac{4\pi}{\sqrt{n_f}}\, \sqrt{\frac{20}{57}} \, \approx \frac{7.4}{\sqrt{n_f}} ~. 
\eeqa
based on the perturbative calculation. For a large coupling $\lambda$ at the scale of $\sim 100$~GeV, the theory could become a strongly coupled one with a nearby Landau-pole scale. The existence of UV fixed point suggests that the UV completion model could have an (approximately) conformal symmetry.

\section{The collider constraints}
\label{sec:collider}

Current collider constraints on this model can be obtained by recasting searches for sleptons at the LHC. In addition to the LHC constraints, $m_\psi$ must be larger than roughly $100~\text{GeV}$ nearly independent of $\Delta m = m_\psi - M_X$ due to constraints from LEP~\cite{Heister:2001nk,Heister:2003zk,Abdallah:2003xe,Achard:2003ge,Abbiendi:2003ji}. The dominant constraints come from standard (uncompressed) slepton searches at CMS~\cite{Sirunyan:2020eab} and ATLAS~\cite{Aad:2019vnb}, as well as from a compressed slepton search at ATLAS~\cite{Aad:2019qnd}. The latter combines a search for missing energy with tagging soft leptons to reduce the QCD background and achieve good sensitivity to lepton portal models for $\Delta m$ between around 1 GeV and 40 GeV, effectively closing the gap for the lepton portal down to 1 GeV splitting and up to a few hundred GeV in mass.

To recast these results, we take the combined observed limits on the selectron and smuon cross section in each search and compare with an leading order calculation of the total electron and muon flavor portal mediator cross section using \texttt{MadGraph5 2.9.2}~\cite{Alwall:2011uj} with a model implemented in \texttt{FeynRules 2.3.41}~\cite{Alloul:2013bka}. These limits are presented by CMS in Figure 14 of Ref.~\cite{Sirunyan:2020eab}, ATLAS in Auxiliary Figure 3a of Ref.~\cite{Aad:2019qnd} for the standard slepton search, and ATLAS in Auxiliary Table 11 of Ref.~\cite{Aad:2019vnb} for the compressed slepton search.  The shaded regions in Fig.~\ref{fig:results} correspond to cross sections exceeding these limits. One can see that the collider results allow a compressed spectrum with the mass splitting below around 1 GeV or between 40 GeV and 100 GeV for $M_X$ below 300 GeV.

\section{Muon anomalous magnetic moment}
\label{sec:muon-g-2}

Loop diagrams with $X$ and $\psi$ can generate a positive contribution to the muon anomalous magnetic moment $a_\mu = (g-2)_\mu/2$.  At one-loop, the new physics contribution from the dark matter sector is
\beqa
\Delta a_\mu^{(X, \psi)} = \frac{n_f\,\lambda^2\,m_\mu^2}{16\pi^2\,M_X^2} \left[ \frac{2+3x- 6 x^2 + x^3 + 6x\ln{x} }{6(1-x)^4} \right] ~,
\eeqa
with $x\equiv m_\psi^2/M_X^2$. In the degenerate limit with $x=1$, the value in the bracket is $1/12$. In the opposite limit with $x\rightarrow 0$ or $m_\psi \gg M_X$, one has
\beqa
\Delta a_\mu^{(X, \psi)} \approx \frac{n_f\,\lambda^2\,m_\mu^2}{16\pi^2\,m_\psi^2} \frac{1}{6} ~. 
\eeqa
Note that the specific combination of $n_f \,\lambda^2$ affects $\Delta a_\mu$. In the degenerate limit and to explain the central experimental value of $\Delta a_\mu$, one needs to have $n_f\,\lambda^2 \approx 4.3\times(M_X/100\,\mbox{GeV})^2$. So, to have a perturbative control of our calculation, we need to consider a lighter $M_X$. Even when $M_X = 100$~GeV, additional two-loop calculations for $\Delta a_\mu$ are needed to the reduce the errors for theoretical predictions. Based on a naive dimension analysis, the two-loop corrections could have a relative error of $\mathcal{O}[n_f\,\lambda^2/(16\pi^2)]$.

\section{Dark matter phenomenology}
\label{sec:dark-matter}

\subsection{Thermal abundance}
\label{sec:thermal}

Although the dark matter abundance could be explained by some non-thermal early universe, the thermal one and especially the weakly interacting massive particle (WIMP) miracle provides a strong motivation for us to correlate the muon $g-2$ measurement to dark matter. So, we discuss the potential parameter space  to have a thermal dark matter in the lepton-portal dark matter model. 

Because $X$ is a real scalar, its self-annihilations to leptons are either helicity suppressed or velocity suppressed. For the leading $s$-, $p$- and $d$-wave ones, the annihilation rate is~\footnote{We use the program \texttt{Calchep}~\cite{Pukhov:2004ca} to calculate some of our formulas.}
\beqa
\label{eq:xx-annih}
 \sigma v(XX \rightarrow \ell^+ \ell^-) = \frac{\lambda^4\,m_{\ell}^2}{4\pi(M_X^2 + m_\psi^2)^2} \, -  \,v^2\, \frac{\lambda^4\,m_{\ell}^2\,M_X^2(M_X^2 + 2 m_\psi^2)}{6\pi\,(M_X^2 + m_\psi^2)^4}  \,+\, v^4\,\frac{\lambda^4\,M_X^6}{60\pi(M_X^2 + m_\psi^2)^4}~,
\eeqa
where for each term we have kept the leading one in $m_\ell^2/M_X^2$. Because of the helicity suppression for both $s$ and $p$ waves, the dominant contribution to the thermal abundance is the $d$-wave one. Defining $\sigma_{\rm eff} v =  \sigma v(XX)=  s + p \,v^2 + d\,v^4$, one has the thermal averaged annihilation rate as $\langle \sigma_{\rm eff} v \rangle = s + 6\,p \, x^{-1} + 60\,d \,x^{-2}$ with $x \equiv M_X/T$. The freeze-out temperature $x_F$ is given by
\beqa
x_F = \ln\left[ \frac{5}{4}\, \sqrt{\frac{45}{8}}\,\frac{g}{2\pi^3} \frac{M_{\rm pl}\,M_X\,(s + 6\,p \, x_F^{-1} + 60\,d \,x_F^{-2})}{\sqrt{g^*}\sqrt{x_F}} \right] ~,
\eeqa
with $g= n_f$ as the number of degrees of freedom for dark matter. Here, $M_{\rm pl} = 1.22\times 10^{19}$~GeV and $g^*$ is the number of relativistic degrees of freedom at $T_F$ and is taken to be 86.25. The dark matter abundance approximately depends on $s$, $p$ and $d$ as 
\beqa
\Omega_X h^2 \approx \frac{1.07\times 10^9}{\mbox{GeV}\,M_{\rm pl}\,\sqrt{g^*}}\, \frac{x_F}{s + 3\,p\,x_F^{-1} + 20\,d\,x_F^{-2}} ~. 
\eeqa
If only one wave dominants, one has $s\approx 0.9\,\mbox{pb}\cdot \mbox{c}$, $p\approx 7\,\mbox{pb}\cdot \mbox{c}$ and $d\approx 27\,\mbox{pb}\cdot \mbox{c}$ to satisfy the observed dark matter abundance. Using \eqref{eq:xx-annih}, one has 
\beqa
d= \frac{\lambda^4\,M_X^6}{60\pi(M_X^2 + m_\psi^2)^4}\, = (27\,\mbox{pb}\cdot \mbox{c}) \times  \left( \frac{\lambda}{3}\right)^4 \,  \left( \frac{100\,\mbox{GeV}}{M_X}\right)^2 \left( \frac{2}{1 +x/4} \right)^{4} ~,
\eeqa
with $x \equiv m_\psi^2/M^2_X$. One needs to have a large coupling $\lambda$ to satisfy the thermal dark matter relic abundance. Otherwise, one may worry about the $X$ abundance is too large and overcloses the universe. We have checked that just using the $d$-wave $XX$ self-annihilation cross section, it is not possible to obtain a thermal dark matter to be consistent with other constraints. On the other hand, additional particles or interactions could be used to make $X$ a thermal dark matter candidate. For instance, one could add an additional Higgs portal coupling to $XX$ to increase the annihilation cross section. 

In the parameter region with an approximately degenerate $X$ and $\psi$ with $\Delta m /M_X \lesssim 1/25$, the coannihilations of dark matter states can provide another interesting thermal dark matter parameter space. Here, we list the formulas for the dominant annihilation processes that have a nonzero $s$-wave annihilation rate. The annihilation rates for $X_a \psi^-_i$ have
\beqa
\sigma v (X_a \psi^-_i \rightarrow \ell_i^- \gamma) = \frac{e^2\,\lambda^2}{32\pi M_X (M_X + m_\psi) } ~, \quad 
\sigma v (X_a \psi^-_i \rightarrow \ell_i^- Z) = \frac{e^2\,\lambda^2\,s_W^2\left[(M_X + m_\psi)^2 - M_Z^2\right]^2}{32\pi\,c_W^2\,M_X (M_X + m_\psi)^5 }  \nonumber ~. 
\eeqa
The flavor-specific two-lepton annihilation rate for $\psi_i^- \psi_i^+$ mediated by $X$ particles is 
\beqa
\sigma v (\psi^-_i  \psi^+_i \rightarrow \ell_i^- \ell^+_i) =  \frac{n_f\,\lambda^4\,m_\psi^2}{32\pi(M_X^2 + m_\psi^2)^2} ~. 
\eeqa
Mediated by photon and the $Z$ boson, one flavor of  $\psi_i^- \psi_i^+$ can also annihilate into all three charged leptons. The rate is given by
\beqa
\sigma v (\psi^-_i  \psi^+_i\rightarrow \gamma^*/ Z \rightarrow \ell^- \ell^+) = \frac{3\,e^4(M_W^4 - 6 M_W^2 m_\psi^2 + 10 m_\psi^4)}{16\pi\,m_\psi^2(M_W^2 - 4 c_W^2 m_\psi^2)^2} ~,
\eeqa
with the overall factor of $3$ from counting three charged leptons. Similarly, the annihilation rates  to quarks  are
\beqa
\sigma v (\psi^-_i  \psi^+_i \rightarrow  \gamma^*/ Z \rightarrow d \overline{d}, s \overline{s}, b \overline{b} ) &=& \frac{3\,e^4(M_W^4 - 2 M_W^2 m_\psi^2 + 10 m_\psi^4)}{48\pi\,m_\psi^2(M_W^2 - 4 c_W^2 m_\psi^2)^2} ~,  \nonumber \\
\sigma v (\psi^-_i  \psi^+_i \rightarrow  \gamma^*/ Z \rightarrow u \overline{u}, c \overline{c}) &=& \frac{2\,e^4(2M_W^4 - 10 M_W^2 m_\psi^2 + 17 m_\psi^4)}{24\pi\,m_\psi^2(M_W^2 - 4 c_W^2 m_\psi^2)^2} ~. 
\eeqa
The annihilation rates to gauge bosons are given by
\beqa
&&\sigma v (\psi^-_i  \psi^+_i \rightarrow \gamma \gamma) = \frac{e^4}{16\pi\,m_\psi^2} ~, \nonumber \\
&&\sigma v (\psi^-_i  \psi^+_i \rightarrow ZZ) = \frac{e^4 s_W^4 (c_W^2 m_\psi^2 - M_W^2)^{3/2} }{4\pi\,c_W^3\,m_\psi (M_W^2 - 2 c_W^2 m_\psi^2)^2} ~, \quad \sigma v (\psi^-_i  \psi^+_i \rightarrow \gamma Z) =  \frac{e^4 s_W^2 (4c_W^2 m_\psi^2 - M_W^2) }{32\pi\,c_W^4\,m_\psi^4} ~,  \nonumber \\
&&\sigma v (\psi^-_i  \psi^+_i \rightarrow W^- W^+) =  \frac{e^4  (4 m_\psi^6 + 16 m_\psi^4 M_W^2 - 17 m_\psi^2 M_W^4 - 3 M_W^6) }{64\pi\,m_\psi^4 (M_W^2 - 2 c_W^2 m_\psi^2)^2} \sqrt{1 - \frac{M_W^2}{m_\psi^2}} ~,
\eeqa
Finally, the annihilation rate to $Z h$ is given by
\beqa
\sigma v (\psi^-_i  \psi^+_i \rightarrow Z h) &=& \frac{e^4\, [16 m_\psi^4 - 8 m_\psi^2(M_h^2 - 5 M_Z^2) + (M_h^2 - M_Z^2)^2] }{1024\pi\,c_W^4\,m_\psi^4 (M_Z^2 - 4 m_\psi^2)^2 }  \nonumber \\
&& \hspace{2.5cm} \times \sqrt{M_h^4 - 2 M_h^2 (4 m_\psi^2 + M_Z^2) + (M_Z^2 - 4 m_\psi^2 )^2} ~. 
\eeqa

In the very degenerate coannihilation region with $\Delta m /M_X \ll x_F$, the effective annihilation rate is given by~\cite{Griest:1990kh}
\beqa
\sigma_{\rm eff} v = \sigma_{ij} \frac{g_i g_j}{g^2_{\rm eff}} = \frac{24}{g_{\rm eff}^2} \left[ \sigma v(\psi^-_i \psi^+_i ) +  n_f \, \sigma v(X_a \psi^-_i ) \right] ~,
\eeqa
by ignoring the small annihilation rate of $\sigma v (XX)$. Here, the total degrees of freedom is $g_{\rm eff} = 12 + n_f$ by including both 3 $\psi$'s and $n_f$ X's.

\subsection{Direct detection}
\label{sec:direct}

\begin{figure}[!htb]
\centering
\includegraphics[width=0.4\textwidth]{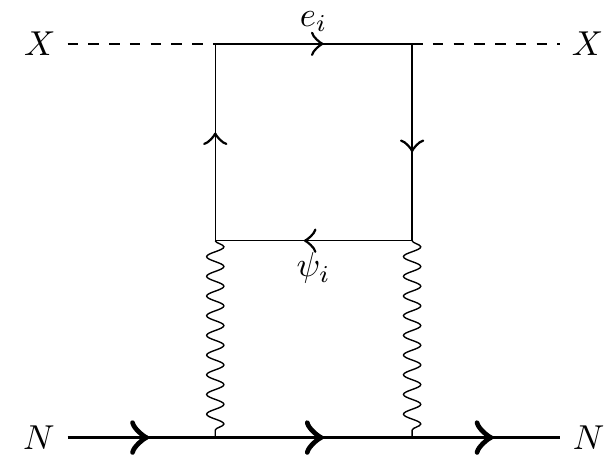}
\caption{Sample diagram for the dark matter candidate scattering off a nucleus at two-loop level. There are several other diagrams corresponding to crossings and interchanging charged-particles.}\label{fig:diag-dd}
\end{figure}
The dark matter scalar $X$ does not have any interactions with hadrons at tree level. At loop level, it couples via photons.  For a real scalar $X$, there is no one-photon coupling, so the leading interaction occurs at two loops and involves the exchange of two photons. A diagram for scattering off a nucleus including an effective two photon operator is shown in Fig.~\ref{fig:diag-dd}. Integrating out the upper box loop in Fig.~\ref{fig:diag-dd}, one has the resulting two photon coupling operator as
\beqa\label{eq:op-dd}
\mathcal{L} \supset -C_1 \,\frac{\lambda^2 \, \alpha}{4 \, \pi \, m_\psi^2} \,  X^2 \, F_{\mu\nu} \, F^{\mu\nu} ~,
\eeqa
with $C_1 = \mathcal{O}(1)$.
This operator generates a coupling to nuclei via the two-photon form factor. Such a coupling has been studied in Ref.~\cite{Pospelov:2000bq}, where it was found that the dark matter-nucleus cross section is given by
\beqa
\sigma_{XN}\,  \approx \, \frac{144 \, \pi}{25} \, \mu_{XN}^2 \, Z^4 \, \alpha^2 \, \frac{\chi_E^2}{r_0^2} ~,
\eeqa
where $\mu_{XN}$ is the reduced mass of the $\chi$-$N$, $Z$ is the atomic number of the nucleus, $r_0$ is the charge radius of the nucleus $r_0 \, \sim \, 1.2~\text{fm} \, A^{1/3}$, and $\chi_{\rm E}$ is the coupling of dark matter to $\mathbf{E}^2$. The latter coefficient is related to the two-photon operator by
\beqa
\chi_{\rm E} \, \sim \, \frac{\lambda^2 \, \alpha}{4 \, \pi \, m_\psi^2 \, M_X} ~.
\eeqa
Spin-independent direct detection constraints are typically presented as the cross section for scattering off a nucleon, which is related to the cross section for scattering off a nucleus by
\beqa
\sigma_{Xn} \, = \, \sigma_{X N} \, \frac{m_n^2}{\mu_{XN}^2 \, A^2} ~,
\eeqa
so that we find for Xenon
\beqa
\sigma_{Xn} \, \sim \, (2 \times 10^{-49}~\text{cm}^2) \, \left(\frac{\lambda}{2.5}\right)^4 \, \left(\frac{150~\text{GeV}}{M_X}\right)^2 \, \left( \frac{150~\text{GeV}}{m_\psi}\right)^4 ~,
\eeqa
which remains out of the reach of current Xenon-based direct detection experiments~\cite{Aprile:2018dbl}.

The scattering cross section of $X$ with a non-relativistic electron can happen at tree-level and is given by  
\beqa
\sigma( X e^- \rightarrow X e^-) \, = \, \frac{\lambda^4\,m_e^4\, (M_X^2 + m_\psi^2)^2}{4\pi M_X^2 (m_\psi^2 - M_X^2)^4}  = (9 \times 10^{-46}~\text{cm}^2) \, \left(\frac{\lambda}{2.5}\right)^4 \, \left(\frac{150~\text{GeV}}{M_X}\right)^2 \, \left( \frac{1~\text{GeV}}{\Delta m }\right)^4
\eeqa
Due to the additional $m_e^2$ suppression of this cross section, this interaction too remains out the reach of current direct detection experiments, which constraint this cross section at the $10^{-38}~\text{cm}^2$ level at $M_X \approx 100~\text{GeV}$~\cite{Aprile:2019xxb}.

\subsection{Indirect detection}
\label{sec:indirect}

The dominant constraint on the lepton portal from indirect detection comes from searches for the peaked tail of internal bremsstahlung photon production. This process produces a photon with a sharp peak just below the maximum photon energy. We calculate the internal bremsstrahlung cross section and compare with the limits on $\sigma v$ obtained in Ref.~\cite{Garny:2013ama} using Fermi-LAT galactic center photon data. The internal bremsstrahlung cross section is given by
\beqa
\langle \sigma v \rangle_\text{IB} \, = \, \frac{\alpha \, \lambda^4}{8 \, \pi^2 \, M_X^2} \, F(m_\psi^2 / M_X^2)\,,
\eeqa
with
\beqa
F(\mu) = (1 + \mu) \, \left[\frac{\pi^2}{6} - \text{log}^2 \frac{1+\mu}{2 \, \mu} - 2 \, \text{Li}_2\left( \frac{1+\mu}{2 \, \mu}\right)\right] + \frac{4 \, \mu + 3}{\mu + 1} + \frac{(4 \, \mu + 1) \, (\mu -1)}{2 \, \mu} \, \text{log}\frac{\mu-1}{\mu+1}.
\eeqa
Note that in $F(1) \to 7/2 - \pi^2/3$ and $F(\infty) \to 4 / (15 \, \mu^4)$.
Our calculation of the cross section agrees with that of Ref.~\cite{Kawamura:2020qxo}. Since this is a line search, it has reduced dependence on astrophysical backgrounds that generally do not yield line-like signatures. Note, however, that these results are subject to uncertainty based on the profile of dark matter in the galactic center, as more peaked profiles lead to regions with enhanced squared density. Profiles including Einasto~\cite{1968PTarO..36..341K}, Navarro-Frenk-White (NFW)~\cite{Navarro:1996gj}, and isothermal vary by as much as a factor of 3 in their projected limits~\cite{Ackermann:2012qk}. For a fixed dark matter mass, as we increase the internal degrees of freedom of dark matter $n_f$, the required coupling $\lambda$ to explain $\Delta a_\mu$ is reduced by a factor of $1/\sqrt{n_f}$. As a result, the indirect detection limit is relaxed by a factor of $n_f^2$. Therefore, we will choose $n_f = 30$ as a representative one to relax the constraints from indirect detection. 

\section{Discussion and conclusions}
\label{sec:discussion}

\begin{figure}[!tbh]
	\centering
	\includegraphics[width=0.47\textwidth]{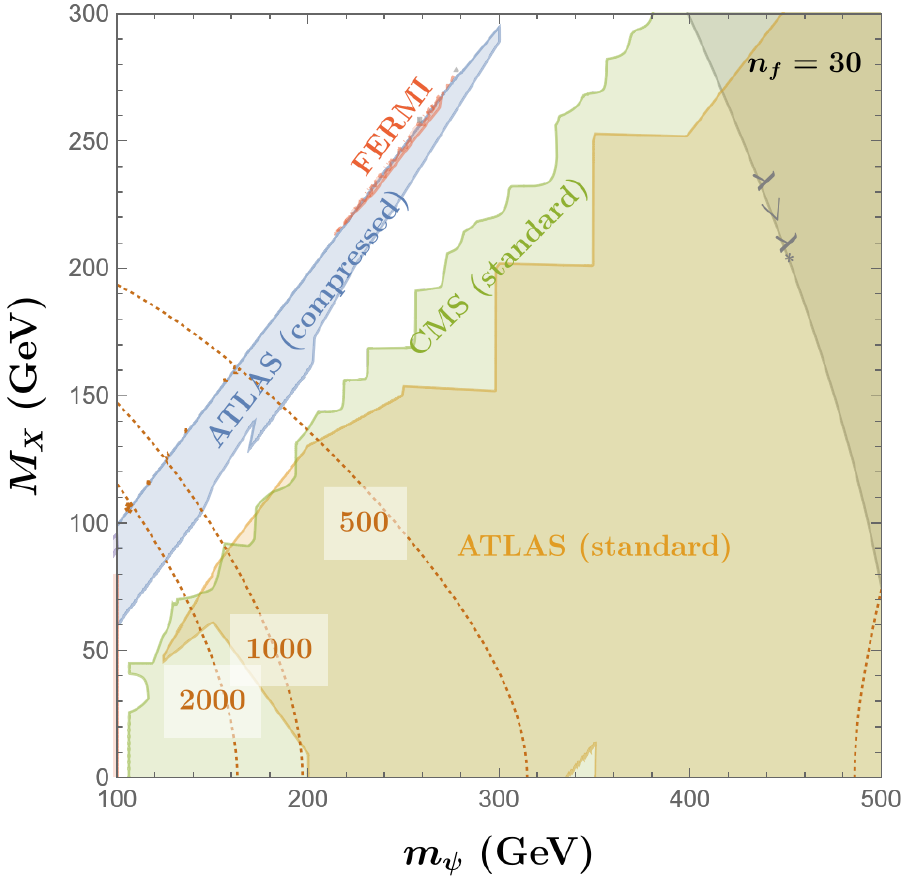}\hspace{0.05\textwidth}\includegraphics[width=0.47\textwidth]{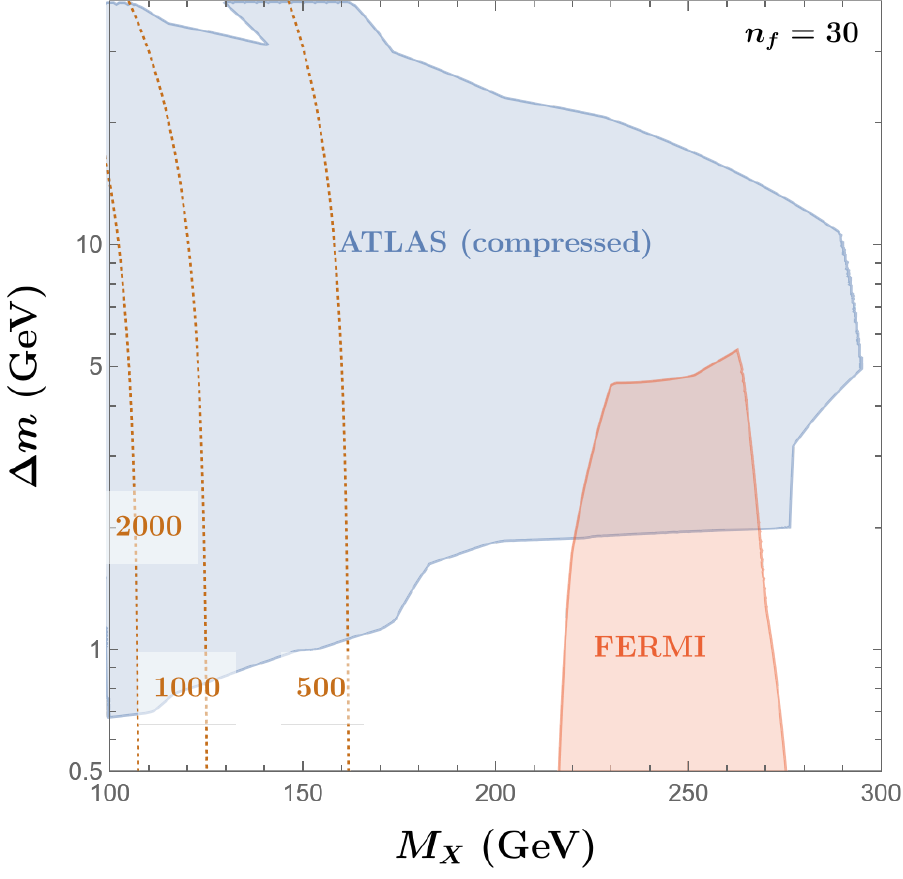}
	\caption{Constraints on the LPDM from collider, indirect detection, and perturbativity after fixing the coupling $\lambda$ to match the central value of the observed $\Delta a_\mu$ and taking $n_f = 30$. The dashed lines indicate the scale in GeV of the Landau pole at which the theory becomes strongly coupled. The left panel shows a broad parameter space, while the right panel focuses on the compressed region.}
	\label{fig:results}
\end{figure}
The results of our study are summarized in Fig.~\ref{fig:results}. In both panels, we fix $\lambda$ to fit the central value of the observed $\Delta a_\mu$. Various constraints from collider and indirect detection as discussed in detail above are shown in the shaded regions. We assume that there are $n_f = 30$ species of dark matter $X$. The shaded gray region indicates the parameter space where the $\lambda$ required to fit $\Delta a_\mu$ is naively greater than $\lambda_*$ of the UV fixed point.  In this region, perturbation theory can certainly not be trusted and higher order corrections will be very important. In addition, due to the relatively large couplings required to fit the observed $\Delta a_\mu$, there will be a nearby Landau pole at which the theory becomes strongly coupled, possibly hitting a conformal fixed point of the $\beta$ function. The scale at which the one-loop running $\lambda(\Lambda) = \lambda_*$ is indicated by the dashed lines for $500~\text{GeV}$, $1~\text{TeV}$ and $2~\text{TeV}$, indicating a Landau pole. The left panel indicates that a $\Delta m$ between roughly 40 GeV and 100 GeV may be accommodated, while the right panel indicates that a $\Delta m$ below 1 GeV may be allowed.

\begin{figure}[!tbh]
	\centering
	\includegraphics[width=0.47\textwidth]{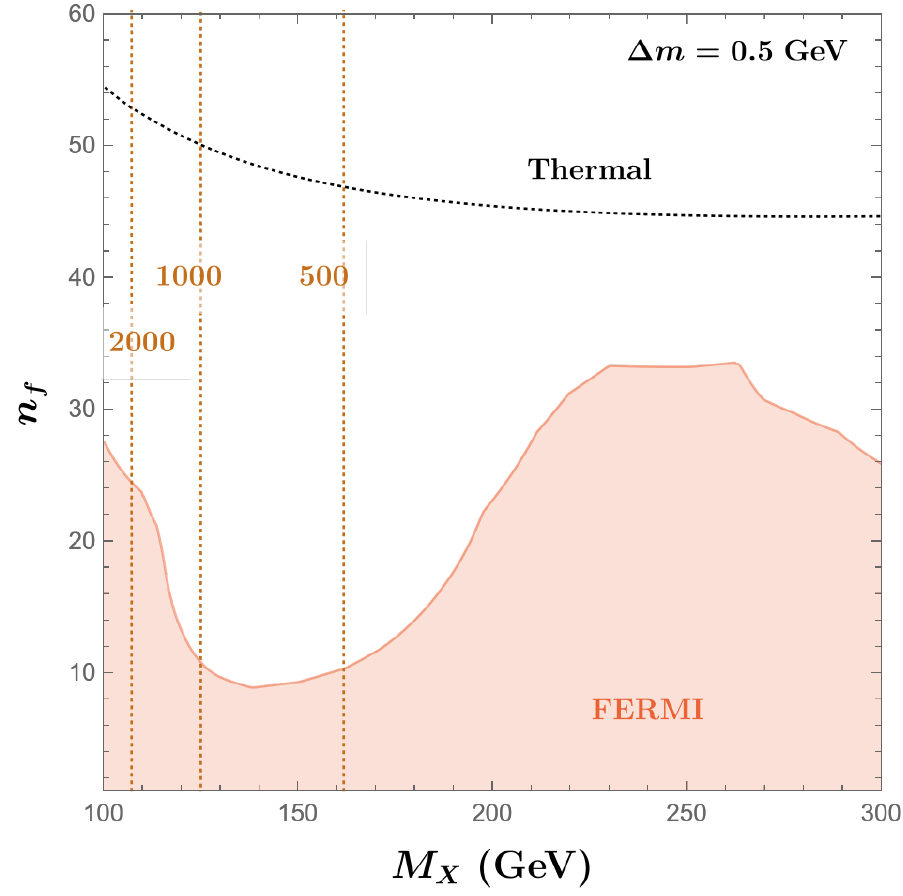}
	\caption{The dark matter mass and $n_f$ for which the model predicts the observed relic abundance of dark matter through co-annihilation. Constraints from Fermi-LAT gamma ray data are also shown in the shaded region, along with the scale of the Landau pole in GeV in the vertical dashed lines.}\label{fig:results-NF}
\end{figure}
It is possible to match the observed abundance of dark matter in a thermal co-annhihilating scenario with large $n_f$ between 45 and 55, with larger $n_f$ preferred in order to maximize the scale of the Landau pole. This behavior is illustrated in Fig.\ \ref{fig:results-NF}. The Fermi-LAT result requires $n_f$ to be larger than around 10, though these constraints are subject to astrophysical uncertainties. 

If such a nearly degenerate scenario is considered, the $\psi_\tau$ is naively stable on collider length scales, as its decay proceeds via an off-shell $\tau$ lepton into a four-body final state. Searches for heavy stable charged particles~\cite{Chatrchyan:2013oca,ATLAS:2014fka} at the LHC rule out this possibility. It is possible, however, to moderately split the $\psi_\tau$ from the $\psi_e$ and $\psi_\mu$ in a minimal flavor violating (MFV)~\cite{Lee:2014rba} scenario, where the non-degenerate correction to the $\psi_i$ mass is proportional to $Y_{e,i}^\dagger\,Y_{e,i}$. Only a small splitting such that $m_\psi - M_X > m_\tau$ is required. The LPDM scenario here bears some resemblance to Flavored Dark Matter~\cite{Agrawal:2011ze,Agrawal:2014ufa}. The flavor structure we consider is simpler in order to avoid constraints from signals like $\mu \to e \gamma$ decay as our focus is on offering an explanation of the Muon $g-2$ result.

In summary, we have found that there exists allowed parameter space in LPDM that explains the $\Delta a_\mu$ excess. The dark matter states including the charged states $\psi_i$ must be below a few hundred GeV. Future experiments can further probe this scenario. At colliders, the high luminosity LHC can improve on the compressed spectrum searches with soft leptons in the final state. A future lepton collider with $\sqrt{s} \gtrsim 500~\text{GeV}$ would nearly completely probe the allowed region. Future gamma ray line searches at Cherenkov Telescope Array~\cite{Actis_2011} and GAMMA-400~\cite{Galper_2013} can also potentially discover the dark matter candidates.

\subsubsection*{Acknowledgements}
The work of YB is supported by the U. S. Department of Energy under the contract DE-SC0017647. The work of JB is supported by start up funds from Colorado State University.

\bibliographystyle{JHEP}
\bibliography{DMmug2}
\end{document}